\documentclass{appolb}
\usepackage{epsfig}

\begin{document}
\title{$e^+e^-$ pair production from nucleon targets in the resonance region%
\thanks{Presented at the XXIX Mazurian Lakes Conference on Physics, Piaski, Poland,
August 30 - Septenber 6, 2005}%
}
\author{Madeleine Soyeur
\address{DAPNIA/SPhN,
CEA/Saclay, F-91191 Gif-sur-Yvette Cedex, France}
\and
Matthias F. M. Lutz
\address{GSI, Planckstrasse 1,  D-64291 Darmstadt, Germany}
}
\maketitle
\begin{abstract}
We discuss two processes, the $\pi\, N \rightarrow e^+e^- \,N$ and
the $\gamma \, N \rightarrow e^+e^- \,N$ reactions, below and close to the
vector meson production threshold ($1.40<\sqrt s <1.75$ GeV). The aim
is to gain understanding of the coupling of vector fields (associated with
$\rho^0$- and $\omega$-mesons) to low-lying baryon re\-sonances. These
couplings are not well-known and important for baryon structure studies
and for dynamical descriptions of vector meson propagation in the
nuclear medium. The  $e^+e^-$ pair production amplitudes
are determined by the $\pi\, N \rightarrow \rho^0 \,N$, $\pi\, N \rightarrow \omega \,N$,
$\gamma \,N\rightarrow \rho^0 N$ and
$\gamma\, N \rightarrow \omega N$ amplitudes supplemented by the Vector Meson Dominance
assumption. The vector meson production amplitudes are calculated consistently using
a relativistic and unitary coupled-channel approach to
meson-nucleon scattering. We display results showing the importance of
the quantum interference between $\rho^0$- and $\omega$-mesons in the $e^+e^-$ channel
to unravel the strength of the coupling of $\rho^0$- and $\omega$-mesons to
specific baryon excitations. The $\pi\, N \rightarrow e^+e^- \,N$ and
$\gamma \, N \rightarrow e^+e^- \,N$ reactions underlie the more complex
 $\pi\, A \rightarrow e^+e^- \, X$ and
$\gamma \, A \rightarrow e^+e^- \,X$ nuclear processes whose measurement is planned
at GSI (with the HADES detector) and under analysis at JLab (with the CLAS detector).
\end{abstract}

\newpage

\section{Introduction}

The $\pi\, N \rightarrow e^+e^- \,N$ and
the $\gamma \, N \rightarrow e^+e^- \,N$ reactions below and close to the
vector meson production threshold ($1.40<\sqrt s <1.75$ GeV) are remar\-kable processes to
study the coupling of low-lying baryon excitations to the $\rho^0$- and $\omega$-meson fields.
The invariant mass of the final time-like photon can be smaller than the physical mass
of the $\rho^0$- and $\omega$-mesons, hence offering the possibility to study the vector
field-nucleon transition couplings of baryon
resonances of masses smaller than 1.72 GeV (the vector meson-nucleon threshold energy).
This possibility relies on the Vector Meson Do\-minance of the electromagnetic
current, an assumption expected to be valid in the vicinity of the vector meson poles \cite{Sakurai,Kroll}.
Another important property of these reactions is the
$\rho^0$-$\omega$ interference in the $e^+e^-$ channel. This effect is large
in the kinematics under consideration and generates patterns in the $e^+e^-$ pair
invariant mass distributions linked to the resonant structure of the amplitudes.
These large interferences can be destructive or constructive, in such a way that
the mere magnitude of the cross sections in a given range of total center of
mass energy provides by itself a strong constraint on the underlying, mostly resonant dynamics.

The interest of our study lies in the availability of detector systems
able to reject the hadronic backgrounds and to measure the small cross sections expected
for dilepton production in $\pi\, N$ and
$\gamma \, N$ reactions. The $\pi\, N \rightarrow e^+e^- \,N$ reaction
could be studied at GSI with the HADES detector and the secondary pion beam \cite{Schoen,HADES}.
The $\gamma \, N \rightarrow e^+e^- \,N$ reaction could be measured with the CLAS
detector at JLab, where inclusive  $e^+e^-$ pair production data on deuteron and nuclear
targets have been taken recently (G7 experiment) \cite{Tur}.

We discuss briefly the processes building up the amplitudes
for the $\pi\, N \rightarrow e^+e^- N$
and $\gamma\, N \rightarrow e^+e^- N$ reactions
and the way they are calculated in Section 2.
We show numerical results for the
$\pi^-\, p \rightarrow e^+e^- n$ and  $\pi^+\, n \rightarrow e^+e^- p$ reactions in Section 3
 and for the $\gamma\, p \rightarrow e^+e^- p$ and $\gamma\, n \rightarrow e^+e^- n$
reactions in Section 4. We conclude by a few comments in Section 5.
The work outlined in this paper relies on two published articles
\cite{Lutz2,Lutz3}.

\section{The $\pi\, N \rightarrow e^+e^- N$
and $\gamma\, N \rightarrow e^+e^- N$ amplitudes}

The graphs entering the calculation of the $\pi\, N \rightarrow e^+e^- N$
and $\gamma\, N \rightarrow e^+e^- N$ amplitudes are displayed in Figs. 1 and 2.
\vglue 0.6truecm
\smallskip
\begin{figure}[h]
\begin{center}
\mbox{\epsfig{file=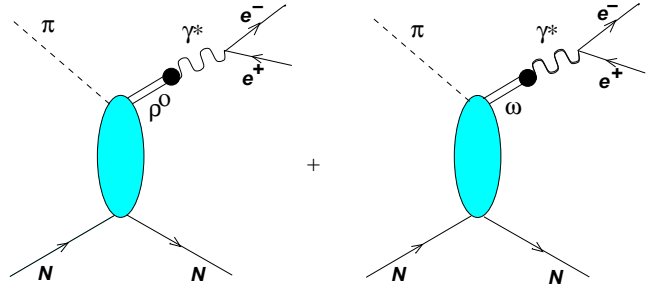, width= 8.5 cm}}
\end{center}
\caption{Diagrams contributing to the amplitude for the $\pi\, N \rightarrow e^+e^- N$ reaction.}
\label{f1}
\end{figure}
\vskip0.9truecm
\begin{figure}[h]
\begin{center}
\mbox{\epsfig{file=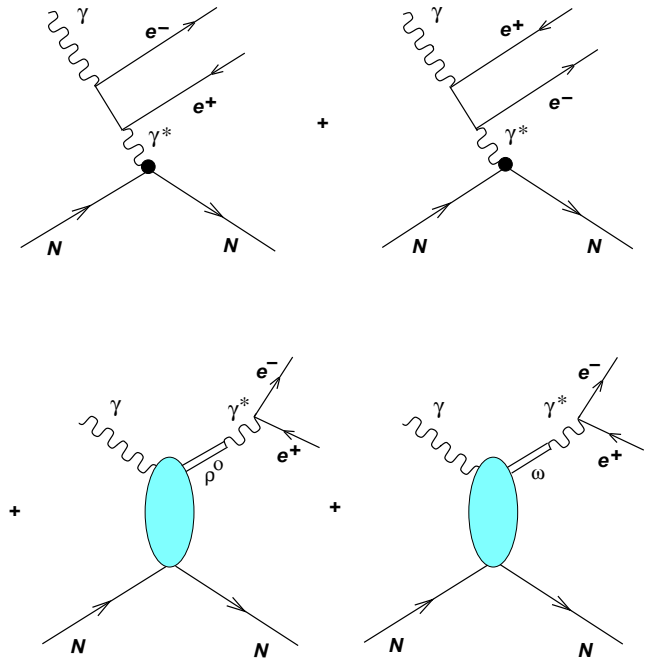, width= 8.5 cm}}
\end{center}
\caption{Diagrams contributing to the amplitude for the $\gamma\, N \rightarrow e^+e^- N$ reaction.}
\label{f2}
\end{figure}

For the $\pi\, N \rightarrow e^+e^- N$ reaction, the amplitude consists of two terms
representing the production of a massive photon of isovector and isoscalar character
respectively. These massive photons
can be related to the $\rho^0$- and $\omega$-meson fields using the Vector Meson Dominance assumption
\cite{Sakurai,Kroll}.
For the $\gamma\, N \rightarrow e^+e^- N$ reaction, there are four terms.
The first two diagrams are associated with the electromagnetic production of
$e^+e^-$ pairs (Bethe-Heitler processes). They depend on the (well-known) nucleon electromagnetic
form factors at low momentum transfers.
The last two diagrams, analogous to the graphs of Fig. 1 for the
$\pi\, N \rightarrow e^+e^- N$ reaction, represent the photoproduction of isovector
and isoscalar massive photons in the Vector Meson Dominance model.

The main dynamical quantities entering the calculation of the $\pi\, N \rightarrow e^+e^- N$
and $\gamma \, N \rightarrow e^+e^- N$ cross sections are therefore the $\pi\, N
\rightarrow \rho N$,
$\pi\, N \rightarrow \omega N$, $\gamma\, N \rightarrow \rho N$ and $\gamma \, N \rightarrow \omega N$
amplitudes. They are computed using the relativistic, chiral coupled-channel approach
of Ref. \cite{Lutz1}. This model offers a consistent picture of the $\pi\, N$ and
$\gamma\, N$ reactions and involves the $\pi N$, $\pi \Delta$, $\rho N$,
$\omega N$, $K \Lambda$, $K \Sigma$ and $\eta N$ hadronic channels. It is restricted to center
of mass energies ranging between $1.40$ GeV and $1.75$ GeV and describes vector meson-nucleon
channels below and very close to the vector meson threshold ($\sqrt s\simeq$1.72 GeV).
The vector meson and the nucleon in the final state are assumed to be
in relative S-wave. The Bethe-Salpeter kernel for the coupled-channel
system is constructed from an effective quasi-local meson-meson-baryon-baryon Lagrangian. The
fundamental fields are the photon, the mesons, the nucleon and the $\Delta(1232)$. The baryon
resonances which do not belong to the large N$_c$ groundstate multiplets are ge\-nerated dynamically \cite{Lutz4}.
They are the N(1520), N(1535), N(1650), $\Delta$(1620) and $\Delta$(1700) resonances.
A generalized Vector Meson Dominance assumption is used to relate amplitudes involving photons to amplitudes
involving vector mesons. The effective Lagrangian parameters are fitted using all available data,
such as phase shifts, inelasticity parameters, pion photoproduction multipole amplitudes, inelastic
pion-nucleon cross sections. The quality of the fit is quite satisfactory for
all these quantities in the interval $1.40$ GeV $\leq \sqrt s \leq 1.75 \,$ GeV \cite{Lutz1}.

The presence of baryon resonances in this energy energy range is
reflected in the structure of the scattering amplitudes for vector
meson production. It is this particular structure that we are interested
in unravelling in $e^+e^-$ pair production processes. We illustrate
the resonant behavior of the vector
meson production amplitudes by showing in Fig. 3 the (projected)
amplitudes for the $\pi\, N \rightarrow \omega N$ process in the
S11 and D13 partial waves \cite{Lutz2}.
In the S${11}$ channel, the N(1535) and the N(1650) resonances
lead to peak structures in the imaginary parts of the amplitudes.
The pion-induced $\omega$ production amplitudes in the D${13}$
channel reflect the strong coupling of the N(1520) re\-sonance to
the $\omega$N channel \cite{Lutz1}. The $\pi N
\rightarrow
\omega N$ amplitudes contain also significant contributions from
non-resonant, background terms. Similar considerations hold
for the other partial waves and in general for the vector meson
production amplitudes in the resonance region \cite{Lutz2,Lutz3}.
\smallskip
\begin{figure}[h]
\begin{center}
\mbox{\epsfig{file=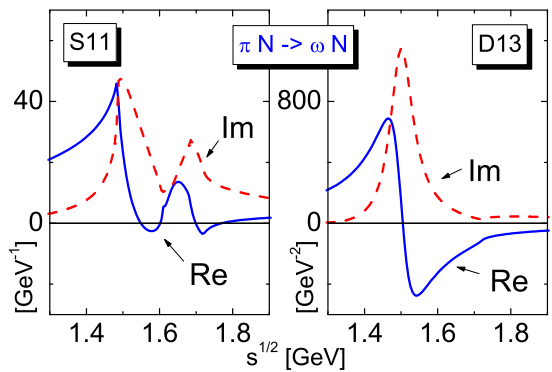, width= 8cm}}
\end{center}
\caption{Amplitudes for the $\pi\, N \rightarrow \omega N$ process in the
S11 and D13 partial waves \cite{Lutz2}.}
\label{f3}
\end{figure}

The $\pi N \rightarrow e^+e^- N$ and $\gamma N \rightarrow e^+e^- N$
amplitudes involving intermediate vector mesons  in the final state are calculated
assuming the specific Vector Meson Do\-minance form given by the
meson-photon interaction terms,
\begin{eqnarray}
{\mathcal L^{int}_{\gamma V}}\,&=&\, \frac {f_\rho} {2 M_\rho^2} F^{\mu \nu}\, \rho^0_{\mu \nu}
\,+\,\frac {f_\omega} {2 M_\omega^2} F^{\mu \nu}\, \omega_{\mu \nu},
\label{eq:e1}
\end{eqnarray}
\noindent
where the photon and vector meson field tensors are defined by
$F^{\mu \nu}\, =\,\partial^\mu A^\nu -\partial^\nu A^\mu$ and
$V^{\mu \nu}\, =\,\partial^\mu V^\nu -\partial^\nu V^\mu.$
This particular form of the Vector Meson Dominance applies
for $e^+e^-$ pairs not too far from the vector meson pole (in contrast
to the generalized Vector Meson Dominance prescription used in Ref. \cite{Lutz1}
needed for extrapolations to the photon point).
In equation (\ref {eq:e1}),
$M_\rho$ and $M_\omega$ are the $\rho$- and $\omega$-masses and
$f_\rho$ and  $f_\omega$ are dimensional coupling constants. Their magnitude can be
determined from the $e^+e^-$ partial decay widths of the $\rho$- and
$\omega$-mesons to be \cite{Friman1}
\begin{eqnarray}
|f_\rho|= 0.036\, GeV^2,
\label{eq:e24}
\end{eqnarray}
\begin{eqnarray}
|f_\omega|= 0.011 \,GeV^2.
\label{eq:e25}
\end{eqnarray}
The relative sign of $f_\rho$ and $f_\omega$ is fixed by
the vector meson photoproduction amplitudes \cite{Lutz1}. We
assume that the phase correlation between isoscalar and isovector
currents is identical for real and virtual photons as in Sakurai's
realization of the Vector Meson Dominance assumption
\cite{Sakurai}. With the conventions used in this paper, both
$f_\rho$ and $f_\omega$ are real and positive.

The Bethe-Heitler terms are computed with phenomenological electromagnetic form factors
\cite{Lutz3}.

Even though we plotted rather broad $e^+e^-$ pair invariant mass spectra to illustrate
general behaviours, we emphasize that our approach
is valid only for values
of m$_{e^+e^-}$ not too far from ($\sqrt s$ - M$_N$). This is a consequence of the
assumption that the vector meson and the nucleon in the final state are
in relative S-wave in the model of Ref. \cite{Lutz1}.

\section{The $\pi\, N \rightarrow e^+e^- N$
reaction: numerical results}
The differential cross sections obtained
for the $\pi^-p \rightarrow e^+e^- n$ and the $\pi^+n \rightarrow
e^+e^- p$ reactions at $\sqrt s$=1.55 GeV are shown in Figs. 4 and
~5. These figures illustrate the overall behaviours of
the invariant $e^+e^-$ invariant mass spectra for these reactions below threshold. For the two
reactions, the $\omega$ and $\rho^0$ contributions to the cross
section are comparable.
This property is quite remarkable in view of the Vector Meson Dominance
coupling constants given in Eqs. (2) and (3) and reflect the
result that the $\pi N \rightarrow \omega N$ amplitudes are much larger than
the $\pi N \rightarrow \rho N$ amplitudes in the model of Ref. \cite{Lutz1}.
The $\rho^0$-$\omega$ interference is
destructive for the $\pi^-p
\rightarrow e^+e^- n$ reaction and constructive for the $\pi^+n
\rightarrow e^+e^- p$ process. Consequently, the $\pi^-p
\rightarrow e^+e^- n$ differential cross section is extremely small
in the range of invariant masses considered in this calculation
(of the order of 10 nb GeV$^{-2}$). In contrast, the constructive
$\rho^0$-$\omega$ interference for the $\pi^+n \rightarrow e^+e^-
p$ reaction leads to a sizeable differential cross section (of the
order of 150 nb GeV$^{-2}$).

This is a very striking prediction,
linked to the resonant structure and magnitude of the scattering
amplitudes. These cross sections reflect indeed the couplings of
both the N(1520) and N(1535) baryon
resonances to the
vector meson-nucleon channels. Data on differential cross sections for the $\pi^-p \rightarrow e^+e^- n$
and $\pi^+n \rightarrow e^+e^- p$ reactions at $\sqrt s$=1.55 GeV would clearly be very
useful for making progress in the understanding of these couplings.

Similar interference patterns are obtained until $\sqrt s$=1.70 GeV.  They are however
sensitive to higher-lying resonances \cite{Lutz2}.

Figs. 6 and
~7 display the $\pi N \rightarrow e^+e^- N$ differential cross section
at $\sqrt s$=1.75 GeV, i.e. right above
the $\omega$-meson
production threshold. The interference pattern changes
dramatically. The differential cross sections for the $\pi^-p
\rightarrow e^+e^- n$ and $\pi^+n\rightarrow e^+e^- p$ reactions
are completely dominated by the $\omega$-contribution. The
magnitudes of the cross sections for the two reactions are now
comparable. The $\rho^0-\omega$ interference is still destructive
in the $\pi^-p \rightarrow e^+e^- n$ channel and constructive in
the $\pi^+n\rightarrow e^+e^- p$ channel, but very small. In
both reactions, crossing the $\omega$-production threshold leads to
a sharp increase in the cross section, by two orders of magnitude
in the $\pi^-p \rightarrow e^+e^- n$ channel and by one order of
magnitude in the $\pi^+n\rightarrow e^+e^- p$ channel.
\begin{figure}[h]
\begin{center}
\mbox{\epsfig{file=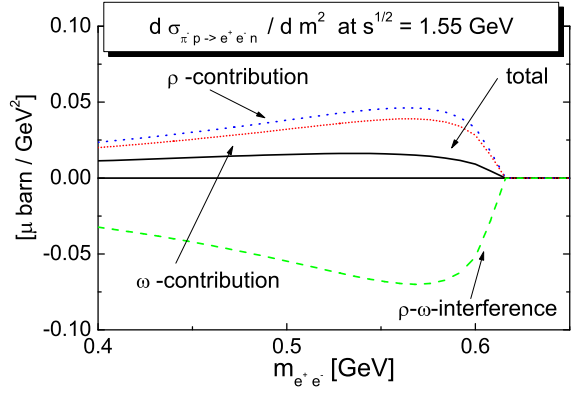, height= 6cm}}
\end{center}
\caption{Differential cross section for the $\pi^-p \rightarrow e^+e^- n$
reaction at $\sqrt s$=1.55 GeV as function of the invariant mass of the $e^+e^-$
pair \cite{Lutz2}.}
\label{f4}
\end{figure}
\smallskip
\begin{figure}[h]
\begin{center}
\mbox{\epsfig{file=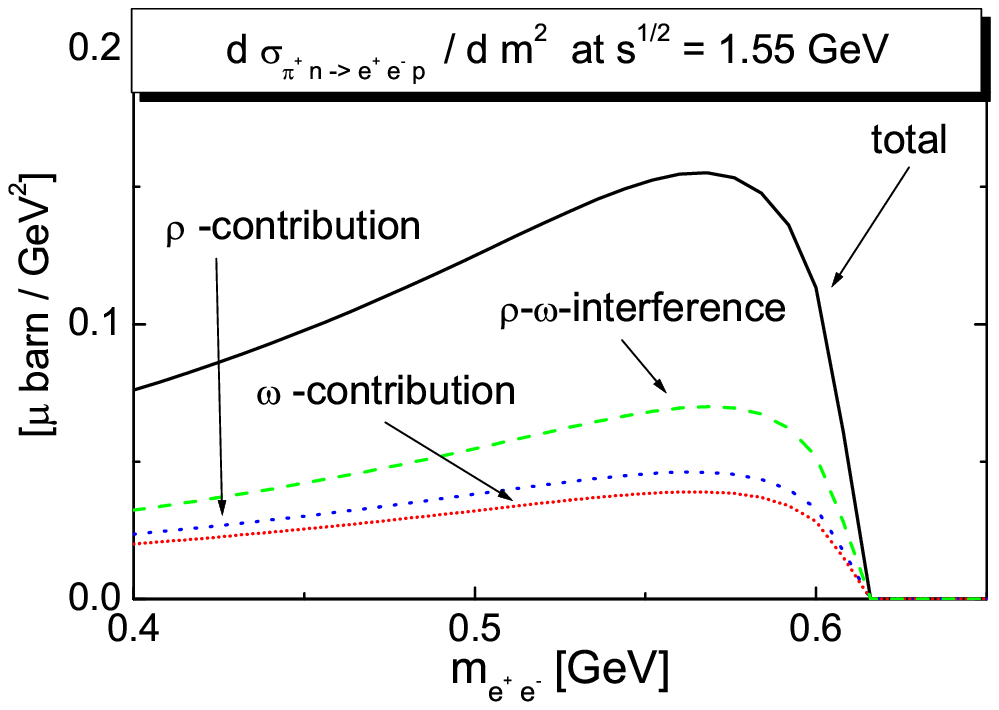, height= 6cm}}
\end{center}
\caption{Differential cross section for the $\pi^+n \rightarrow e^+e^- p$
reaction at $\sqrt s$=1.55 GeV as function of the invariant mass of the $e^+e^-$
pair \cite{Lutz2}.}
\label{f5}
\end{figure}
\smallskip
\begin{figure}[h]
\begin{center}
\mbox{\epsfig{file=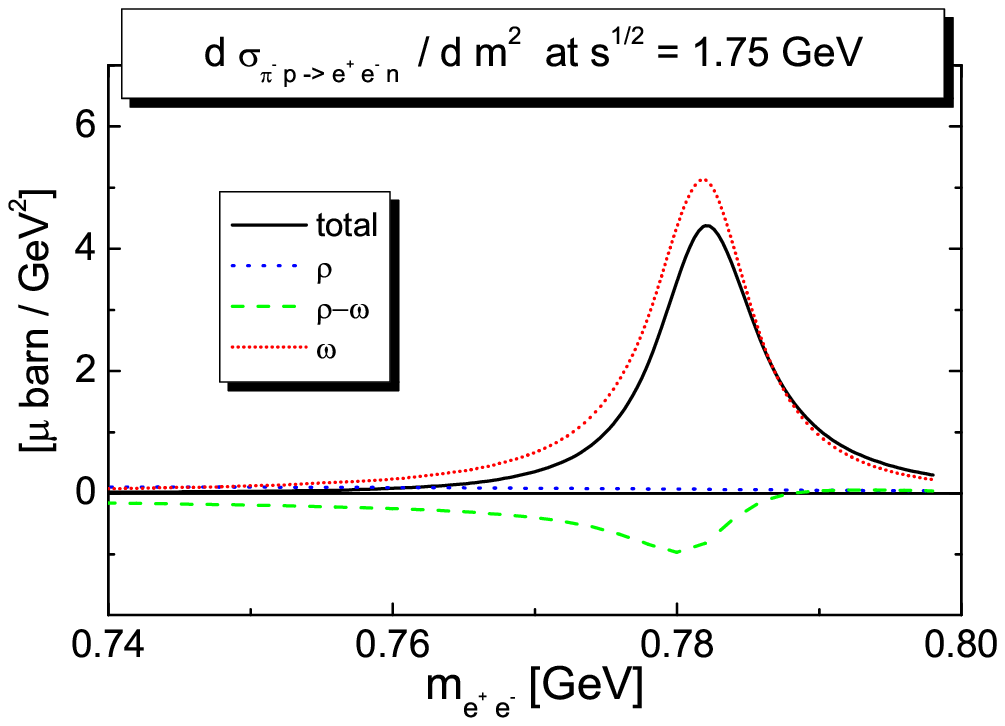, height= 6cm}}
\end{center}
\caption{Differential cross section for the $\pi^-p \rightarrow e^+e^- n$
reaction at $\sqrt s$=1.75 GeV as function of the invariant mass of the $e^+e^-$
pair \cite{Lutz2}.}
\label{f6}
\end{figure}
\smallskip
\begin{figure}[h]
\begin{center}
\mbox{\epsfig{file=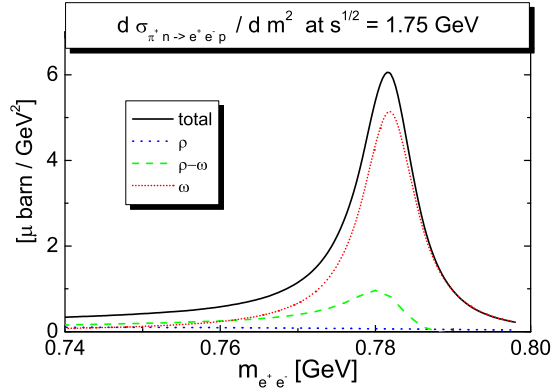, height= 6cm}}
\end{center}
\caption{Differential cross section for the $\pi^+n \rightarrow e^+e^- p$
reaction at $\sqrt s$=1.75 GeV as function of the invariant mass of the $e^+e^-$
pair \cite{Lutz2}.}
\label{f7}
\end{figure}
\smallskip

We note that the general interference pattern of the $\pi^-p \rightarrow e^+e^- n$ and
$\pi^+n \rightarrow e^+e^- p$ reactions found in our calculation has also been obtained in Ref. \cite{Titov},
with rather different absolute magnitudes of the cross sections. The latter result is linked
to a very different prescription for the transition couplings of baryon resonances to
the $\rho$-nucleon and $\omega$-nucleon channels.

\section{The $\gamma\, N \rightarrow e^+e^- N$
reaction: numerical results}
We show first $e^+e^-$
pair spectra where the leptonic phase space is fully integrated. There is no interference between
Bethe-Heitler and vector meson decay
amplitudes in that case \cite{Lutz3}. Such results are presented in Fig. 8
for the $\gamma\,p \rightarrow e^+e^- p$ and
$\gamma\,n \rightarrow e^+e^- n$ reactions at $\sqrt s$=1.55 GeV. To unravel the dynamics of the
dilepton production process, we display separately the Bethe-Heitler and vector meson decay
contributions to the cross sections and the decomposition of the vector meson decay
cross sections into the $\rho$, $\omega$ and $\rho$-$\omega$ interference terms for the two possible
spin channels (J=1/2 and J=3/2).
\begin{figure}[h]
\mbox{\epsfig{file=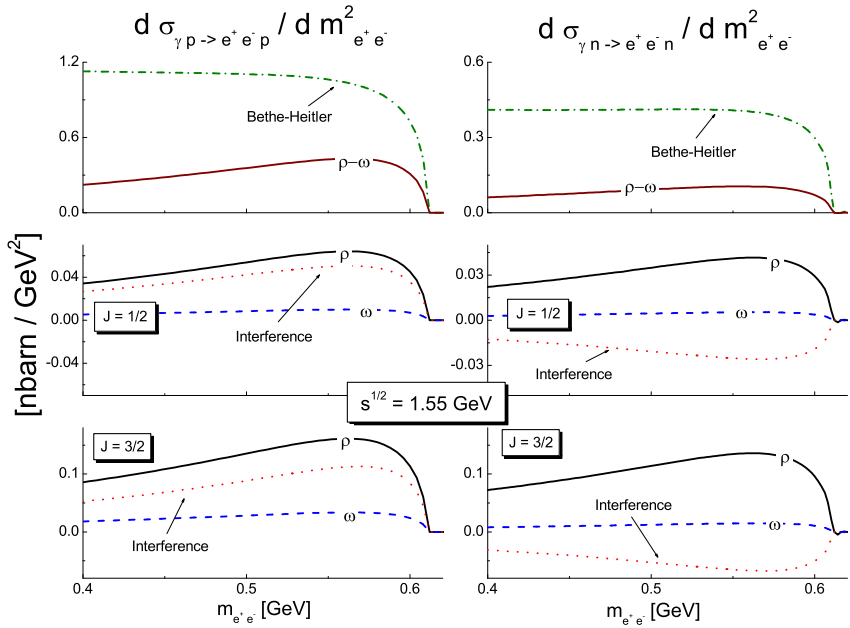, height= 9.5cm}}
\caption{ Integrated spectra for the $\gamma\,p \rightarrow e^+e^- p$ and
$\gamma\,n \rightarrow e^+e^- n$ reactions
at $\sqrt s$=1.55 GeV together with their different components \cite{Lutz3}.}
\label{f8}
\end{figure}
We notice first that the Bethe-Heitler process is the dominant
contribution to the differential cross sections. All experimental studies
of the vector meson photoproduction amplitudes below threshold in the
dilepton channel will therefore require a careful subtraction of the Bethe-Heitler contribution.
The second observation is that the $\rho$-meson production and decay
into $e^+e^-$ pairs dominates the vector meson contribution.
This is expected in view of the Vector Meson Dominance
coupling constants given in Eqs. (2) and (3), if no particular dynamical effect
enhances the $\omega$-meson contribution (as in the $\pi\, N \rightarrow e^+e^- N$ reaction).
The $\gamma\,N \rightarrow e^+e^- N$ reaction is therefore a very useful process
to study $\rho$-meson photoproduction below threshold.
Thirdly we found that the quantum interference between $\rho$- and $\omega$-meson ${e^+e^-}$ decays
is constructive for proton targets and destructive for neutron targets,
in both J=1/2 and J=3/2 channels.
The effects of this interference are however less pronounced than in the
case of the $\pi\, N \rightarrow e^+e^- N$ reaction because of the relative
smallness of the $\omega$-meson contribution.

The interference of Bethe-Heitler pairs with $e^+e^-$ pairs produced by the decay
of vector mesons reflects in asymmetries \cite{Korchin2,Lvov2}.
They are defined by subdividing the lepton pair phase space
into two hemispheres. Considering first the rest frame of the produced $e^+e^-$ pair,
we call forward and backward electrons those characterized by $\cos\, (\vec p_-, \vec q\,) > 0$
and  $\cos\, (\vec p_-, \vec q\,) < 0$ respectively,
where $q\,,p_+\,$ and $p_-$ are the photon, positron and electron momenta. If the $e^+e^-$ pair is moving,
we can project the cross section onto the forward and backward hemispheres using the
$\theta$ functions, $\theta\,[+ q\,(p_+ - p_-)]$ and $\theta\,[- q\,(p_+ - p_-)]$
\cite{Lutz3}.
We refer to these projections as $\sigma^+$
and $\sigma^-$. The symmetric and asymmetric part of the differential cross sections are
given by
\begin{eqnarray}
\biggl[\frac{d\sigma^{sym}}{d{m_{e^+e^-}}^2 }\biggr]_{\gamma N \rightarrow e^+e^- N} =
\biggl[\frac{d\sigma^+ }{d{m_{e^+e^-}}^2}\biggr]_{\gamma N \rightarrow e^+e^- N}
+ \biggl[\frac{d\sigma^-}{d{m_{e^+e^-}}^2}\biggr]_{\gamma N \rightarrow e^+e^- N},
\label{def:sigsym}
\end{eqnarray}
\begin{eqnarray}
\biggl[\frac{d\sigma^{asym}}{d{m_{e^+e^-}}^2 }\biggr]_{\gamma N \rightarrow e^+e^- N} =
\biggl[\frac{d\sigma^+ }{d{m_{e^+e^-}}^2}\biggr]_{\gamma N \rightarrow e^+e^- N}
- \biggl[\frac{d\sigma^-}{d{m_{e^+e^-}}^2}\biggr]_{\gamma N \rightarrow e^+e^- N}.
\label{def:sigasym}
\end{eqnarray}
We restrict our discussion to the results obtained for the
$\gamma\,p \rightarrow e^+e^- p$ reaction
at $\sqrt s$=1.65 and 1.75 GeV, displayed in Fig. 9. We refer to \cite{Lutz3} for a much more extended
presentation.
\begin{figure}[h]
\begin{center}
\mbox{\epsfig{file=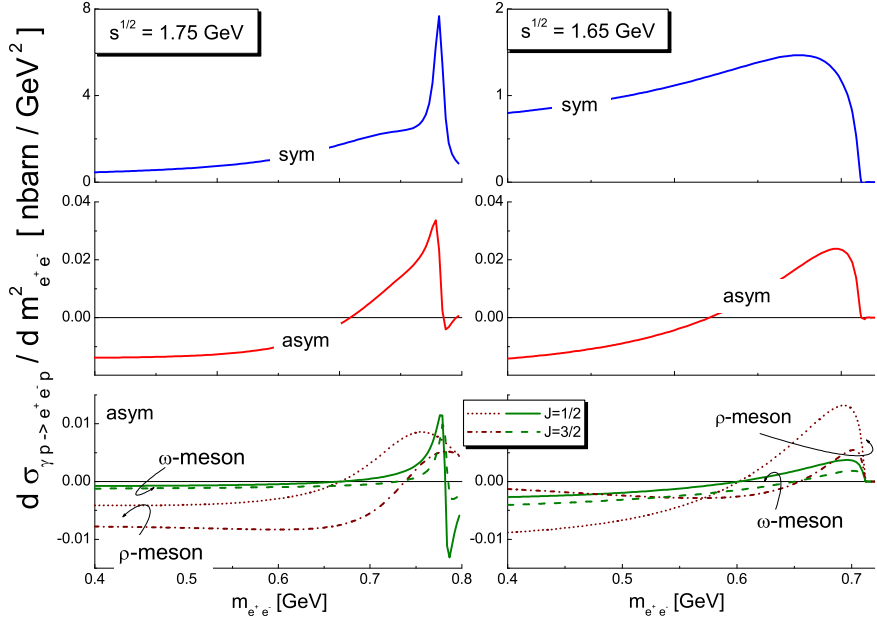, height= 9.5cm}}
\end{center}
\caption{Symmetric and asymmetric contributions to the $\gamma\,p \rightarrow e^+e^- p$ cross section
at $\sqrt s$=1.75 and 1.65 GeV together with their different components \cite{Lutz3}.}
\label{f9}
\end{figure}
The symmetric part is the total cross section and hence
the sum of the Bethe-Heitler and vector meson cross sections. The antisymmetric part reflects
the interference of the Bethe-Heitler and vector meson pairs.
The comparison of the symmetric and antisymmetric cross sections
shows that the antisymmetric cross section is
much smaller (by more than two orders of magnitude in the
mass range of interest) than the symmetric cross section.
The asymmetric cross section consists of terms reflecting the interference
of Bethe-Heitler pairs with $\rho$-meson and $\omega$-meson decay pairs respectively.
We see from Fig. 9 that the Bethe-Heitler $\rho$-meson interference
is the dominant contribution, except for pairs arising from
the decay of $\omega$-mesons produced on the mass-shell slightly above threshold.

Angular distributions for the different subprocesses indicate that the pairs originating from
Bethe-Heitler processes and vector meson decays are produced in very
different regions of phase-space \cite{Lutz3}. The Bethe-Heitler pairs are emitted at
forward angles while
$e^+e^-$ pairs from vector meson decays are produced isotropically in
the center of mass.
The Bethe-Heitler spectra peak strongly at low $e^+e^-$ pair invariant masses,
while vector meson decay is enhanced by the proximity of the poles for large $e^+e^-$ pair masses.
Consequently the quantum interference between the two processes is very small
and most significant at small angles where the Bethe-Heitler cross section
is very large.

Our results suggest that the vector meson contribution can be determined quite accurately from
experimental $e^+e^-$ spectra by subtracting the
Bethe-Heitler term and neglecting the small
interference
of Bethe-Heitler pairs with vector meson $e^+e^-$ decays.

\section{Conclusion}

We have calculated consistently the $\pi \,N \rightarrow e^+e^- N$ and $\gamma\,N \rightarrow e^+e^- N$
cross sections below and just above
the vector meson production threshold ($1.40<\sqrt s <1.75$ GeV).
The $\rho$- and $\omega$-meson production amplitudes underlying these processes are computed
using the relativistic and unitary coupled-channel approach
to meson-nucleon scattering of Ref. \cite{Lutz1}, supplemented
with the Vector Dominance assumption (1) for the time-like
photon in the final state.

We have found that the $\pi \,N \rightarrow e^+e^- N$ and $\gamma\,N \rightarrow e^+e^- N$
reactions are unique and complementary processes to study the couplings of selected
baryon resonances to the vector meson-nucleon channels. The complementarity
comes from the fact that the $\pi^-p \rightarrow e^+e^- n$ and the $\pi^+n \rightarrow
e^+e^- p$ reactions are very
sensitive to the couplings of selected baryon resonances to both $\rho^0$- and $\omega$-mesons
in our model while
the $\gamma\,N \rightarrow e^+e^- N$ process is ideally suited to study
the coupling of low-lying baryon resonances to the $\rho^0$-nucleon channel.
The quantum interference of $\rho^0$- and $\omega$-mesons in the $e^+e^-$ channel
amplifies the sensitivity of the $\pi \,N \rightarrow e^+e^- N$ and $\gamma\,N \rightarrow e^+e^- N$
reactions to the coupling of vector fields to baryon resonances.

Experimental data on both processes in the kinematics under consi\-deration
are expected in the near future, from the CLAS
detector at JLab for the $\gamma \, N \rightarrow e^+e^- \,N$ reaction \cite{Tur} and
from the HADES detector and the secondary pion beam at GSI for the $\pi\, N \rightarrow e^+e^- \,N$ reaction
\cite{Schoen,HADES}.

\vskip 1 true cm
\noindent
{\bf Acknowledgement}

One of us (M.S.) is grateful to the Organizers of the XXIX Mazurian Lakes Conference on
Physics for inviting her to a most pleasant and enri\-ching meeting.

\end{document}